\documentclass[lettersize,journal]{IEEEtran}
\usepackage{graphicx} 
\usepackage{amsmath,amssymb,amsfonts}
\usepackage{braket}
\usepackage{xcolor}
\usepackage{hyperref}
\usepackage[caption=false, font=scriptsize]{subfig}
\usepackage{blindtext}
\usepackage{tikz}
\usepackage{nomencl}
\makenomenclature
\usepackage{orcidlink}

\usepackage{cite}

\usepackage{upgreek}

\DeclareMathOperator*{\argmin}{arg\,min}
\DeclareMathOperator*{\argmax}{arg\,max}

\title{Evaluating Quantum Optimization for Dynamic Self-Reliant Community Detection}
\author{David Bucher\orcidlink{0009-0002-0764-9606}, Daniel Porawski\orcidlink{0000-0002-7286-505X}, Benedikt Wimmer\orcidlink{0009-0004-5481-594X}, Jonas Nüßlein\orcidlink{0000-0001-7129-1237}, Corey O'Meara\orcidlink{0000-0001-7056-7545
}, Naeimeh Mohseni\orcidlink{0000-0003-3373-4572}, Giorgio Cortiana\orcidlink{0000-0001-8745-5021} and Claudia Linnhoff-Popien
\thanks{D. Bucher, D. Porawski, and B. Wimmer are with Aqarios GmbH, Prinzregentenstraße 120, 81677 Munich, Germany, e-mail: \href{mailto:david.bucher@aqarios.com}{david.bucher@aqarios.com}.}%
\thanks{J. Nüßlein and C. Linnhoff-Popien are with Ludwig-Maximilians University Munich, Institute for Computer Science, Mobile and Distributed Systems Chair, e-mail: jonas.nuesslein@ifi.lmu.de.}%
\thanks{C. O'Meara, N. Mohseni, and G. Cortiana are with E.ON Digital Technology GmbH, Laatzener Str. 1, 30539 Hannover, Germany, e-mail: \{corey.o’meara, giorgio.cortiana\}@eon.com.}%
}

\begin{document}
\maketitle
\begin{abstract}
Power grid partitioning is an important requirement for resilient distribution grids. Since electricity production is progressively shifted to the distribution side, dynamic identification of self-reliant grid subsets becomes crucial for operation.
This problem can be represented as a modification to the well-known NP-hard Community Detection (CD) problem. We formulate it as a Quadratic Unconstrained Binary Optimization (QUBO) problem suitable for solving using quantum computation, which is expected to find better-quality partitions faster. The formulation aims to find communities with maximal self-sufficiency and minimal power flowing between them. To assess quantum optimization for sizeable problems, we apply a hierarchical divisive method that solves sub-problem QUBOs to perform grid bisections.
Furthermore, we propose a customization of the Louvain heuristic that includes self-reliance. In the evaluation, we first demonstrate that this problem examines exponential runtime scaling classically. Then, using different IEEE power system test cases, we benchmark the solution quality for multiple approaches: D-Wave's hybrid quantum-classical solvers, classical heuristics, and a branch-and-bound solver. As a result, we observe that the hybrid solvers provide very promising results, both with and without the divisive algorithm, regarding solution quality achieved within a given time frame.
Directly utilizing D-Wave's Quantum Annealing (QA) hardware shows inferior partitioning.
\end{abstract}


\bstctlcite{IEEEexample:BSTcontrol}


\nomenclature[01]{$N$}{Number of nodes in the grid}
\nomenclature[02]{$\mathcal{C}$}{Community structure: Partition of nodes}
\nomenclature[03]{$N_P$}{Number of producers in the grid}
\nomenclature[04]{$K$}{Number of possible communities}
\nomenclature[05]{$A$}{Adjacency matrix with absolute powerflow entries}
\nomenclature[06]{$w_{i,j}$}{Power flow on the line connecting $i$ and $j$}
\nomenclature[07]{$k_i$}{Sum of absolute power flowing from and to node $i$}
\nomenclature[08]{$m$}{Sum of total absolute power flow}
\nomenclature[09]{$p_i$}{Power consumption/production at node $i$}

\nomenclature[10]{$x_{i,\ell}$}{Binary variable: 1 iff node $i$ in comunity $\ell$, else 0}
\nomenclature[11]{$Q(\mathcal{C})$}{Modularity for $\mathcal{C}$}

\nomenclature[12]{$P(\mathcal{C})$}{Net power deviation}
\nomenclature[13]{$E(\mathcal{C})$}{Self-reliant community energy}
\nomenclature[14]{$\mathcal{P}$}{Normalizing constant}
\nomenclature[15]{$\rho$}{One-hot constraint penalty factor}
\nomenclature[16]{$\lambda$}{Self-Reliance penalty parameter}

\nomenclature[17]{$e^{\text{rel}}$}{Relative solution quality}
\nomenclature[18]{$B$}{Number of lines between two communities}

\printnomenclature

\section{Introduction}\label{sec:introduction}
The ever-growing integration of renewable electricity in the grid causes a shift from a centrally planned operation of the electricity system to a decentralized, local one. This trend is expected to continue in the future to a state where potentially all electricity can be produced locally~\cite{colucci2023}. In such a future electricity grid, the planning periods are expected to be considerably shorter since the production side is intermittent and large loads like charging Electric Vehicles (EV) or heat pumps are dynamically switched on and off.

The separation of the grid into virtual microgrids aids analysis and administrative tasks~\cite{cotilla-sanchez2013}, and positively affects energy balancing~\cite{koukaras2022} as well as storage utilization~\cite{chen2014}. Especially with growing decentral electricity production, logical clusters may help in incentivizing the prosumer behavior of individual customers because interaction does not happen between customers and the grid operator but between customers within the same community~\cite{fernandez-campoamor2021b}.

This paper proposes an extension to the commonly known combinatorial optimization problem for grid partitioning called Community Detection (CD): The Self-Reliant Community Detection (SRCD) problem finds self-sufficient virtual microgrids, i.e., partitions of the producers and consumers, where production and consumption are almost balanced. 
This is achieved by including the momentary power flow in the community detection formulation and adding a second self-reliance objective that aims to minimize the mismatch between production and consumption.
The found community structure is not expected to be fixed for one network but is expected to change over time with different consumption and production data of the customers. E.g. community structure may shift when an EV is plugged in at one part of the grid while it another one finished charging. The grid operator can use such dynamic identification of the self-reliant partitions to implement dynamic pricing zones or reward self-sufficient customers with better tariffs. Alternatively, P2P markets can be established within the community, effectively 
reducing the administrative overhead of matching consumer-producer pairs in self-sustaining grids~\cite{omeara2023}. This comes with the benefit of trading electricity, where it physically flows.

CD based on modularity maximization is an NP-hard combinatorial problem~\cite{brandes2006}. Solving CD classically is typically done with heuristics, like Girvan-Newman~\cite{pahwa2013} or Louvains~\cite{blondel2008} algorithm, leading to good approximate solutions. Yet, solving the problem to optimality requires potentially exponential runtime, which we show in Sec.~\ref{sec:evaluation}. Since SRCD has to be optimized on a regular basis, fast and good solutions are desired.

In recent years, Quantum Computing (QC) has gained attention since the ever-improving hardware now allows for the first proof-of-concept applications~\cite{abbas2023}. Quantum optimization is one of the prominent application areas of QC, where it is able to efficiently find solution of NP-hard combinatorila optimization problems faster than classical algorithms. Solving SRCD with quantum optimization can lead to identifying a new quantum utility use case, i.e., an optimization problem where quantum optimization techniques deliver good quality solutions practically faster or equally as fast as classical algorithms.

The expectation of QC to speed up optimization tasks stems from the inherent quantum mechanical properties, superposition, entanglement, and an exponentially growing state space. Grover's algorithm~\cite{grover1996} is one of the first quantum algorithms that exhibits a scaling advantage compared to classical search algorithms. Although equipped with a mathematically proven speedup, the practical use of Grover is still not possible due to the Noisy Intermediate Scale Quantum (NISQ) state of current days hardware and the relatively deep circuits involved~\cite{wang2020}.
Instead, short, variational circuits, like the Quantum Approximate Optimization Ansatz (QAOA)~\cite{farhi2014, mastroianni2024a} or the Variational Quantum Eigensolver (VQE), are the focus of current-day research on quantum optimization applications, with the hope of tapping into quantum utility soon. Besides optimization algorithms on universal quantum computers, purpose-built Quantum Annealing (QA) devices for optimization have emerged~\cite{farhi2001, rajak2023}. They rely on the adiabatic principle of quantum mechanics that states if the evolution of a system happens slowly enough, the system stays in the ground state of the momentary Hamiltonian~\cite{born1928, farhi2001}.  In that sense, a QA device is initialized with an easily preparable ground state. Subsequently, the system undergoes a gradual transition to a Hamiltonian, where the ground state of the system encodes the solution to a specific optimization problem. Currently, available QA hardware is provided by the company D-Wave and features the usage of quadratically interacting Ising problem Hamiltonians~\cite{mcgeochb}. This problem specification is isomorphic to the Quadratic Unconstrained Binary Optimization (QUBO) problem, in which many NP-complete and NP-hard problems can be formulated~\cite{lucas2014b}. The application of QA in various domain-specific application tasks has been widely explored in literature~\cite{feld2019, phillipson2021, omeara2023, fernandez-campoamor2021b, bucher2023, colucci2023}. CD has also been studied using QA ansätze~\cite{ushijima-mwesigwa2017a, reittu2019, gemeinhardt2021a, stein2023}.


Since the goal of the work is to demonstrate the current performance of QA and QA-enhanced techniques, we formulate the SRCD as a QUBO. However, since current hardware does not allow for solving large QUBO instances because of restricted connectivity, we employ a hierarchical divisive hybrid quantum algorithm that recursively splits the network using solutions of smaller QUBO sub-problems. Similar methods have already successfully been applied for solving related problems~\cite{gcs-q, ushijima-mwesigwa2017a}.
Finally, we benchmark the approaches on various test cases with IEEE power system test case topology. As a classical baseline, we utilize the Louvain algorithm that needs to be modified to include power flow modularity and the self-reliance penalty, as well as Simulated Annealing (SA)~\cite{kirkpatrick1983} and QBsolv~\cite{booth2017a} as a classical QUBO heuristics, and Gurobi~\cite{gurobi} as a classical branch-and-bound solver. Additionally, we utilize D-Wave's hybrid cloud solvers and their quantum annealing computer as hybrid and quantum methods.
The results obtained directly from D-Wave's processor are considerably worse than all other solving strategies, despite successful hyperparameter tuning. This is due the the noise being persistent in today's unmatured hardware. The modified Louvain algorithm, along with the proprietary hybrid solvers from D-Wave, performs best with respect to relative solution quality achieved within a given time window.


The main contributions of this paper are as follows:
\begin{enumerate}
    \item The novel formulation of the SRCD problem including its QUBO-formulation.
    \item Customization of the Louvain algorithm and the implementation of the divisive method for solving the SRCD.
    \item Extensive benchmarking and evaluation on different test cases to demonstrate the competitiveness of quantum approaches.
\end{enumerate}


The remainder of the letter is structured as follows: We overview related work on graph partitioning and quantum annealing applications in Sec.~\ref{sec:related_work}. Afterwards, we introduce Quantum Annealing and discuss the classical and hybrid methods used in Sec.~\ref{sec:background}. Then, we present our new mathematical formulation of the SRCD model in Sec.~\ref{sec:formulation}, followed by the problem-specific solution concepts in Sec.~\ref{sec:methods}. The subsequent Sec.~\ref{sec:evaluation} explains the methodology employed for evaluating the performance of the solvers, and Sec.~\ref{sec:results} presents the results of the benchmarks. Finally, we conclude our findings in Sec.~\ref{sec:conclusion}.

\section{Related Work}\label{sec:related_work}

\subsection{Grid Partitioning}
Operating power grids is a challenging and complex task that becomes increasingly complicated when integrating intermittent renewable energy resources. Grid partitioning has emerged as a potential tool to overcome these issues and facilitate operation, control, and management~\cite{zhao2019}. Physical partitions can help in making the system more resilient~\cite{li2010} and efficient by eliminating cascading effects~\cite{jia2017}, loop flows, and unnecessary losses~\cite{kamwa2007}. Furthermore, separation can minimize costs as the number of new lines to be constructed can be reduced~\cite{colucci2023}.

Grid partitioning based on operational experience or administrative regions is limited due to the inaccurate representation of the underlying structure~\cite{zhao2019}. As a result, algorithmic methods have emerged as more effective alternatives~\cite{kamwa2007}. Graph theoretic ansätze based on hierarchical spectral clustering have been explored, featuring grid-specific properties like admittance and average power flow~\cite{sanchez-garcia2014}. Furthermore, Ref.~\cite{cotilla-sanchez2013} explores partitioning through k-means clustering using an electrical distance, paying attention to cluster size and count.

Besides the mentioned methods, grid partitioning through CD has been widely explored in literature~\cite{guerrero2018, xu2019, pahwa2013}. CD can be approximately solved using the methods from before (k-means and spectral clustering)~\cite{fortunato2010, xu2019} or with heuristic algorithms (Girvan-Newman~\cite{pahwa2013}, Louvain~\cite{blondel2008}). Moreover, genetic algorithms have been identified as encouraging solution strategies for modularity-based grid partitioning~\cite{guerrero2018}. However, since modularity solely considers the topology of the underlying grid and leaves out its functioning, adaptations of the modularity have been made to identify so-called functional community structures based on Electrical Coupling Strength (ECS)~\cite{zhao2019, wang2022}.
The CD methods listed here are entirely static, in opposition to SRCD, which is a dynamic approach that includes the momentary power draw and infeed of the customers in the power grid. The idea of self-reliance is not directly applicable to the static CD approaches.

\subsection{Applications of Quantum Optimization}
Quantum optimization has been applied to many industry-relevant topics of different domains by transforming known optimization problems into QUBO~\cite{lucas2014b}. Application areas include logistics, where QA has been successfully employed to solve the capacitated vehicle routing problem~\cite{feld2019}, which is an extension of the renowned traveling salesman problem. In finance, QA has been used to optimize the composition of a portfolio of assets using Markovitzian portfolio theory~\cite{phillipson2021}, which maps naturally into QUBO form. Electricity domain use cases include the application of QAOA to solve the prosumer problem, i.e., efficiently scheduling loads given a dynamic tariff to minimize overall costs of consumers~\cite{mastroianni2024a}, or a hybrid QA approach to schedule tariff discounts in order to achieve demand side response and improve utilization of renewable electricity~\cite{bucher2023}.

Grid partitioning with different objectives has also successfully been achieved with QA: 
Ref.~\cite{fernandez-campoamor2021b} explore the application of QA to community detection on IEEE grid topologies with ECS in the modularity, like Ref.~\cite{zhao2019} introduced.
Ref.~\cite{wang2023} investigates grid partitioning based on the Laplacian matrix in the system. To formulate the inequality constraint containing problem as QUBO they introduce approximate integer slack variables and observe faster runtime of the QA-based method. Both of the grid partitioning methods are static as opposed to our dynamic approach. Ref.~\cite{colucci2023} explore hybrid QA grid partitioning with the goal of minimizing the cost for constructing microgrids while enabling electricity sharing~\cite{tanjo2016a}. For that they formulate the problem as QUBO with integer slack variables likewise to the method proposed in Ref.~\cite{wang2023}. However, even though they include self-reliance constraint, it is terms of electricity surplus throughout one year of entire regions. Therefore, their formulation is not directly applicable to our intended goals.


\section{Background}\label{sec:background}


\subsection{Quantum Annealing}
In quantum mechanics, the elementary unit of information, a quantum bit (or qubit), can be in the superposition of both basis configurations $\ket{\psi} = \alpha \ket{0} + \beta \ket{1}$, with $|\alpha|^2 + |\beta|^2 = 1$. Combining $n$ qubits together allows for encoding all $2^n$ bit configurations in the quantum state, i.e.,
\begin{align}
    \ket{\psi} = \sum_{x = 0}^{2^n - 1} \psi_x \ket{x},
\end{align}
where $x \in \{0, 1\}^n$. Measuring the quantum state in the computational basis reveals the bit string $x$ with probability $|\psi_x|^2$~\cite{nielsen2010}. Thus, the goal of quantum optimization lies in preparing a quantum state with a high probability of measuring the optimal bit configuration regarding a combinatorial optimization problem.

Utilizing the Pauli $z$-matrix $\hat \sigma^z_i = \ket{0}\bra{0} - \ket{1}\bra{1}$ acting on qubit $i$, one can define an Ising Hamiltonian (cost operator)
\begin{align}\label{eq:ising_model}
    \hat H_{C} = -\sum_{i, j} J_{i,j} \hat \sigma^z_i \otimes \hat \sigma^z_j -\sum_{i} h_i \hat \sigma^z_i,
\end{align}
where the tensor product of the Pauli operator refers to the operator acting on qubits $i$ and $j$ simultaneously. The Ising Hamiltonian is diagonal in the $z$-basis. Therefore, its ground state is a computational basis state, meaning a bit string, now referred to as $x^*$.
The energy (or cost) of a bit string can also be described entirely classically, i.e., $\bra{x}\hat H_C \ket{x} = H_C(x)$, where $H_C$ is of QUBO form
\begin{align}
    H_C(x) = -\sum_{i,j}J_{i,j} z_i z_j  - \sum_i h_i z_i = \sum_{i,j} Q_{i,j} x_i x_j,
\end{align}
with $z_i = 1 - 2 x_i$, or $\hat \sigma^z_i \ket{x_i} = z_i \ket{x_i}$. Therefore, once it is possible to encode a combinatorial problem as QUBO, one can implement the Ising Hamiltonian physically and solve the problem with the following method~\cite{lucas2014b}.

The adiabatic theorem of quantum mechanics states that if a system starts in the ground state of a Hamiltonian and that Hamiltonian begins to vary slowly, it remains in the ground state of the momentary Hamiltonian~\cite{born1928, farhi2001}. As a consequence, Adiabatic Quantum Computing (AQC) begins in the equal superposition state $\ket{+} = 2^{-n/2}\sum_x \ket{x}$, the ground state of $\hat H_D = - \sum_i \hat \sigma^x_i$, and undergoes the time evolution of the following Hamiltonian
\begin{align}\label{eq:time-evolution}
    \hat H(s) = s \hat H_C + (1 - s) \hat H_D,
\end{align}
with $s \in [0, 1]$. When the transition from $s = 0$ to $s = 1$ happens slowly enough, the state will transition to the ground state of $\hat H_C$, i.e., $\ket{x^*}$, which is the solution to the QUBO encoded in $\hat H_C$.

Due to decoherence and other noise artifacts, the transition between the two Hamiltonians can never happen slowly enough in reality. Therefore, one has to decrease the transition time, leaving the AQC regime of the process and entering Quantum Annealing (QA)~\cite{rajak2023}. Nevertheless, the probability of measuring the optimal bit string $|\psi_{x^*}|^2$ normally does not vanish. So, repeatedly sampling from the resulting probability distribution may eventually reveal the optimal solution.

D-Wave's current QA hardware, or Quantum Processing Unit (QPU), the Advantage System\footnote{In our case, Advantage 5.4, located in Germany.} offers well above 5000 qubits and more than 35000 couplers~\cite{mcgeochb}. The qubits are arranged in the so-called Pegasus topology, connecting every qubit to 15 other qubits. The fixed topology poses a practical caveat when the QUBO connectivity is larger than 15 because multiple physical qubits must be tied together to form a logical qubit (called \emph{chain}) with higher connectivity. This reduces the size of a fully connected QUBO that can be solved using the Advantage QPU to only 177 binary variables.

\subsection{Classical and Hybrid Methods}
To solve optimization problems classically (or hybrid), we will utilize the following methods throughout the paper:

Gurobi~\cite{gurobi} is one of the most advanced Mixed-Integer-Programming (MIP) solvers commercially available. At its core, it is a branch-and-bound optimizer that features a wide variety of constraints and quadratic objective functions.

SA~\cite{kirkpatrick1983} is a Monte Carlo (MC) method for solving QUBO problems. In each MC sweep, SA proposes bit flips in its current solution and accepts them according to the energy of the configuration and an ever-decreasing temperature. Thus, SA gradually transitions from exploring the solution space to exploiting a (local) minimum. 

QBsolv~\cite{booth2017a} is a hybrid solution strategy initially created for solving large instances with QA. It creates and solves sub-problem QUBOs by fixing certain bits in the problem and only optimizes for those with the largest energy impact. In conjunction, a classical Tabu search~\cite{glover1986} is performed on the complete problem. Both solvers are wrapped in an outer loop. As a sub-problem solver, we utilize SA in our experiments to remain entirely classical. QBsolv with SA will be called QSA in the rest of the text.
 
D-Wave's Leap Cloud platform offers proprietary hybrid quantum-classical solvers that harness both the power of their QA hardware as well as classical computational power. Since the QPU devices are too small (or too sparsely connected) for direct execution of large problems, they internally perform decomposition to map problems onto the hardware. The Binary Quadratic Model (BQM) solver (called \emph{Leap BQM}) is capable of solving QUBOs while the Constrained Quadratic Model (CQM) solver (called \emph{Leap CQM}) can handle constraints and integer/real variables.

\section{Mathematical Formulation}\label{sec:formulation}

\begin{figure*}
    \centering
    \input{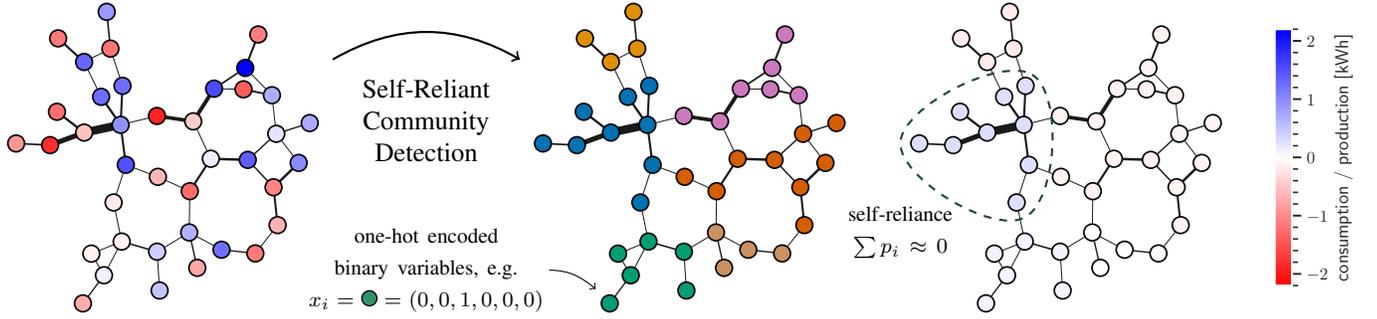}
    \caption{Examplatory instance of a power grid with producers and consumers (\textsf{case39}). Negative values refer to production, and positive values refer to consumption. The line thickness indicates the absolute power flowing through that line. The community of each node is described through the one-hot encoding of $K$ binary variables. The net consumption within a community is aimed to be minimized to guarantee self-reliance.}
    \label{fig:case39-example}
\end{figure*}

Given a power grid as a graph $G(V,E)$ with $|V| = N$, we want to find a mutually exclusive community structure $\mathcal{C} = \{C_1, \dots, C_K\}$, with $C_i \subseteq \{1,\dots,N\}$, $C_i \cap C_j = \emptyset \,\forall i \neq j$, and $\bigcup_i C_i = \{1,\dots,N\}$. Each node represents a consumer or producer in the grid with $p_i$ being its momentary load, if positive, and production, if negative. The production and consumption data are considered pre-existing for the graph partitioning problem and will not be varied throughout the optimization. 
The community structure $\mathcal{C}$ should be optimized such that most of the electricity is flowing within nodes of the same partition, and the total generation within the community closely matches the total consumption.
Therefore, in this section, we construct a self-reliant version of the modularity-based CD, the so-called Self-Reliant Community Detection (SRCD). We account for the first property by executing CD on a graph with edge weights resembling the physical power flow. We include the second property through a self-reliance penalty term in the optimization objective. 
Lastly, the problem must be translated into a QUBO for quantum optimization.
Fig.~\ref{fig:case39-example} shows an exemplary instance of the SRCD problem and its desired solution. 

\subsection{Preprocessing}\label{sec:preprocessing}
First of all, we need to compute the physical power flow for the given consumption and production data of the nodes $p_i$. To this end, we perform a linearized power flow computation using the python package \texttt{pandapower}~\cite{pandapower2018}. Despite linear DC power flow not being the ideal choice for distribution-style power grids, we still utilize it due to its simplicity. The core focus of this paper is the performance of QC techniques on this kind of problem and not the most realistic power flow values. The power flow computation can be replaced by more advanced methods without issues.

As a result of the DC power flow, we get the power flowing over the line connecting $i$ with~$j$ as $w_{i,j} \in \mathbb{R}$. We, therefore, construct a graph with the edge weights being the absolute power flow over the respective line, described by the adjacency matrix
\begin{align}\label{eq:pf-adj}
    A_{i,j} = \begin{cases}
    |w_{i,j}| \quad&\text{if } (i, j) \in E\\
    0 &\text{else}.
    \end{cases}
\end{align}

\subsection{Modularity}
Modularity is a measure of the community structure of a graph partition. High modularity means the community structure given by $\mathcal{C}$ matches the intrinsic partitions of the graph well. The modularity can take values between $[-1/2, 1]$ and is defined as follows~\cite{brandes2008}:
\begin{align}
\label{eq:modularity}
    Q(\mathcal{C}) = \frac{1}{2m}\sum_{C\in \mathcal{C}} \sum_{i,j \in C}\left( A_{i,j} - \frac{k_i k_j}{2m}\right),
\end{align}
where $k_i = \sum_j A_{i,j}$ and $m = \frac{1}{2}\sum_i k_i$. Consequently, finding a $\argmax_\mathcal{C} Q(\mathcal{C})$ that maximizes the modularity will uncover grid partitions. Hence, this is the main objective of our optimization problem. Powerflow modularity, i.e. using the actual power flow as the connection strength within the grid topology from Eq.~\eqref{eq:pf-adj}, helps identify self-reliant communities. Little power is flowing between approximately self-reliant network partitions as all of the produced electricity is consumed within.

\subsection{Self-Reliance}
As the second part, we want to make sure that the power produced within a community also gets consumed within it. This condition is most likely not exactly enforceable, but only approximately, i.e., $\sum_{i \in C_j} p_i \approx 0\, \forall j$, as we cannot control the loads $p_i$ in our setting. 
We can quantify the squared \emph{net power deviation} from zero for all communities combined as follows
\begin{align} \label{eq:self-reliance}
   P(\mathcal{C}) = \frac{1}{\mathcal{P}}\sum_{C \in \mathcal{C}} \left( \sum_{i \in C} p_i \right)^2,
\end{align}
with $\mathcal{P} = \sum_i p_i^2$ being a normalization constant that helps with combining the objective with the modularity. Minimizing $P(\mathcal{C})$ finds customer groupings with minimal net power deviation. Instead of the absolute power deviation, the squared power deviation can be naturally mapped onto a QUBO problem, as the next section shows.

\subsection{QUBO Formulation}
\label{sec:qubo-formulation}
To eventually formulate the problem as QUBO, we must represent the community structure $\mathcal{C}$ through binary variables.
To this end, we must first restrict ourselves to a maximum of $K$ possible communities. Then, we use one-hot encoding: Each node is associated with $K$ bits, of which only a single one is set to 1, indicating the community index the node has been assigned to. An example of one-hot encoding can be seen in Fig.~\ref{fig:case39-example}. The total number of bits is $N \times K$. Using the notation $x_{i,\ell} \in \{0, 1\}$, i.e. $x_{i,\ell} = 1$ indicates node $i$ belongs to community $\ell$, we formulate the following identity
\begin{align}\label{eq:one-hot-delta}
    \sum_{\ell = 1}^K x_{i,\ell} x_{j,\ell} = \begin{cases}
        1 \quad&\text{if $i$ and $j$ belong to}\\&\text{the same community}\\
        0 &\text{otherwise.}
    \end{cases}
\end{align}

Additionally, the one-hot encoding has to be enforced by making sure the following condition is satisfied
\begin{align}\label{eq:one-hot-constraint}
   &\sum_{\ell = 1}^{K}x_{i,\ell} = 1 \, \forall i
   \quad\Longleftrightarrow\quad \min_x \sum_i \left(\sum_{\ell=1}^K x_{i,\ell} - 1\right)^2.
\end{align}
The right-hand expression of the equivalence is the penalty term required for QUBO~\cite{lucas2014b}.

The number of communities $K$ is generally unknown, but a too-large $K$ is not an issue because these bits do not have to be assigned. A problem only arises if $K$ is chosen to be smaller than the real optimal number of communities. In our experiments, we utilize a heuristic approach to determining a good value of $K$, which is described in Sec.~\ref{sec:srcd-experiments}.

Combining community detection~\eqref{eq:modularity} with the self-reliance penalty~\eqref{eq:self-reliance}, we arrive at the following minimization objective
\begin{align}
\label{eq:optimization}
    \mathcal{C}^* = \argmin_\mathcal{C}\left[ -Q(\mathcal{C}) + \lambda P(\mathcal{C}) \right],
\end{align}
i.e., maximizing modularity while minimizing deviations from the self-reliance condition. Here, $\lambda > 0$ is a parameter that lets us tune the strength of the self-reliance penalty.
In Sec.~\ref{sec:evaluation}, we will discuss how we set $\lambda$ in our experiments.

Using the binary variables and the one-hot encoding we subsequently get $\sum_{i \in C_\ell} p_i = \sum_{i = 1}^{N} x_{i\ell} p_i$. Therefore, together with Eq.~\eqref{eq:one-hot-delta}, we can formulate Eq.~\eqref{eq:optimization} into a QUBO problem, with $X =\{0, 1\}^{N \times K}$:
\begin{gather}
    \min_{x \in X} \Bigg[-\frac{1}{2m}\sum_{i,j} \left(A_{i,j} - \frac{k_i k_j}{2m} \right) \sum_{\ell} x_{i,\ell} x_{j,\ell} \nonumber\\+ \frac{\lambda}{\mathcal{P}} \sum_\ell \left( \sum_i p_i x_{i,\ell} \right)^2 + \rho \sum_i \left(1 - \sum_\ell x_{i,\ell}\right)^2 \Bigg ],
\label{eq:srcd-qubo}\end{gather}
where the $\rho$ term is the penalty from Eq.~\eqref{eq:one-hot-constraint} that enforces the one-hot constraint. The penalty factor $\rho > 0$ has to be chosen very large for the solution to be valid. When solving the problem with MIP solvers (such as Gurobi or Leap CQM), one can directly specify the constraint Eq.~\eqref{eq:one-hot-constraint} without adding the penalty term.

%
%
%
\subsection{Consumption Bias}

In real data $\sum_i p_i \neq 0$, i.e., there is not equally as much power generated in the overall grid as consumed. Thus, one can include $K' < K$ communities that are excluded from the self-reliance penalty~\eqref{eq:self-reliance}. For example, if most of the nodes consume electricity, only a few self-sustaining microgrids can be found. The remaining customers are gathered in the $K'$ non-self-reliant microgrids that consume all their power from an external grid. In the upcoming experiments, we will only consider completely self-reliant test cases.

\section{Custom Optimization Algorithms}\label{sec:methods}
In this section, we present two problem-specific methods for solving the SRCD problem. One is a derivative of the well-known Louvain algorithm for CD, a classical heuristic that agglomoratively identifies communities. Second, we propose a divisive method that iteratively splits the grid into smaller sub-grids until no further improvement is achievable. The splitting problem has QUBO form and can, therefore, be solved by all QUBO solvers mentioned before, including QA. This makes the divisive algorithm effectively a hybrid quantum-classical method for solving SRCD.

\subsection{Louvain algorithm}
Several greedy heuristics for solving the graph partitioning problem based on modularity maximization exist~\cite{clauset2004c, blondel2008, traag2019}. One of the most widely used algorithms is called the \emph{Louvain}-method~\cite{blondel2008}. It works by agglomeratively merging communities in order to find an optimal community structure. In contrast to the QUBO~\eqref{eq:srcd-qubo}, the algorithm does not require the total number of communities predetermined. In the following, we develop an adaptation to Louvain that includes the self-reliance condition in its formulation.

Initially, Louvain starts with all nodes in the network assigned to their own communities. It picks a node at random and computes the modularity differences $\Delta Q$ for moving the selected node into the communities of all of its neighbors. The move with the largest modularity gain will be picked and executed. Afterwards, a new node is selected and the procedure is repeated until no positive modularity gain is achievable anymore. Instead of computing the difference by computing the entire modularity, $\Delta Q$ can be directly computed, making Louvain very efficient. The modularity difference of moving node $i$ out of community $D$ and into community $C$ can be computed as follows
\begin{align}
    \Delta Q(D \rightarrow i \rightarrow C) = \Delta Q(D \rightarrow i) + \Delta Q(i \rightarrow C)
\end{align}
with, moving $i$ into community $C$ being expressed as
\begin{gather}
    \Delta Q(i \rightarrow C) = \left[\frac{k_C^\text{in} + 2 k_{i,C}}{2m} - \left(\frac{k^\text{tot}_C + k_i}{2m}\right)^2\right] \nonumber\\
    - \left[\frac{k_C^\text{in}}{2m} - \left(\frac{k^\text{tot}_C}{2m}\right)^2 - \left(\frac{k_i}{2m}\right)^2\right] = \frac{k_{i,C}}{m} - \frac{k_i k_C^\text{tot}}{2m^2}.
\end{gather}
Here, $k_{i,C} = \sum_{j \in C} A_{i,j}$ are all the edge weights between $C$ and $i$, $k_C^\text{in} = \sum_{i,j \in C} A_{i,j}$ the edge weights within $C$ and $k_C^\text{tot} = \sum_{i \in C} k_i$ all the edge weights of the community members. The expression for the opposite move $\Delta Q(D \rightarrow i)$ works similarly~\cite{blondel2008}.

We need to include another term into the modularity difference that accounts for the self-reliance condition:
\begin{align}
    &\Delta P(i \rightarrow C) \nonumber\\&= \frac{1}{\mathcal{P}} \left[\left(p_i + \sum_{j \in C} p_j\right)^2 -\left(\sum_{j \in C} p_j\right)^2 - p_i^2 \right] \\
    &= \frac{2p_i}{\mathcal{P}} \sum_{j \in C} p_j.
\end{align}
Likewise, $\Delta P(D \rightarrow i) = -\frac{2p_i}{\mathcal{P}} \sum_{j \in D, i \neq j} p_j$.

The central quantity of the Louvain method can then be adapted as follows
\begin{align}
    &\Delta \tilde{Q}(D \rightarrow i \rightarrow C) \nonumber
    = \Delta Q(D \rightarrow i \rightarrow C)\\ &- \lambda\frac{2p_i}{\mathcal{P}} \left(\sum_{j \in C} p_j - \sum_{k \in D, k \neq i} p_k \right).
\end{align}
All remaining parts of the algorithm do not need to be modified.

The complete Louvain method encloses the procedure described to this point with an outer loop that coarse grains the graph after each step, i.e. merging all found communities in a single node. Then, the procedure described above is repeated on the coarse-grained graph, and the outer loop terminates if modularity gain can no longer be achieved. For more detail, see Ref.~\cite{blondel2008}.

\subsection{Divisive Algorithm}
The ordinary CD problem has already been solved by using a recursive bipartition strategy~\cite{ushijima-mwesigwa2017a}. Additionally, divisive methods have successfully solved Coalition Structures Generation (CSG) problems in induced subgraph games through the GCS-Q algorithm~\cite{gcs-q, venkatesh2023b}. Therefore, inspired by these methods, we derive the optimal splitting sub-problem QUBO for SRCD in the following, starting from the full problem formulation in Eq.~\eqref{eq:srcd-qubo}.
The divisive algorithm works by recursively finding bipartite splits and dividing the problem into multiple sub-problems. In contrast to Louvain, it begins with all nodes in the same community and terminates when no further split can improve the solution quality.
Performing only bipartite splits lets us drop the one-hot-encoding constraint and represent the community assignment by a single binary variable: $y_i=0$ indicates node $i$ belongs to the first community and $y_i=1$ means it belongs to the second one for the particular sub-problem being solved. With respect to the formulation of the single-problem approach \eqref{eq:srcd-qubo}, the QUBO problem for a sub-problem can be formulated as follows: Let $S$ be the set of nodes in the sub-problem to solve, with
$Y =\{0, 1\}^{\vert S\vert}$ and $S\subseteq \{1,\dots,N\}$.
We can utilize $\delta_{y_i, y_j} = 2y_i y_j - y_i - y_j + 1$ as the Kronecker delta of two categories. Hence, the QUBO formulation for the sub-problem reads
\begin{gather}
    \min_{y \in Y} \Bigg[-\frac{1}{2m}\sum_{i,j} \left(A_{i,j} - \frac{k_i k_j}{2m} \right)\delta_{y_i, y_j}\nonumber\\
    +\frac{\lambda}{\mathcal{P}} \left(\sum_ip_i y_i \right)^2 +\frac{\lambda}{\mathcal{P}} \left(\sum_ip_i (1-y_i) \right)^2  \Bigg ],
\label{eq:divisive-qubo}\end{gather}
effectively reducing the size of the QUBO matrix of the first bipartite split by the factor of $K$ compared to the single-problem QUBO. Furthermore, each subsequent problem is roughly half the size of the problem from which it was derived. Note that QUBO~\eqref{eq:divisive-qubo} is equivalent to the Min-Cut QUBO found in GCS-Q~\cite{gcs-q}.

Small sub-problem QUBOs can often be solved reliably and fast using classical methods. We, therefore, extend Refs.~\cite{ushijima-mwesigwa2017a, gcs-q} by employing a cut-off size for which problems smaller than the size will be solved classically. This will alleviate any fine branching towards the end, which would lead to many QPU calls. Furthermore, since the SRCD sub-problem QUBOs are fully connected, we employ fast Clique embedding~\cite{boothby2016} as the strategy to map the problem to the D-Wave hardware graph, effectively reducing wall clock times.

Notably, the divisive approach is greedy, and a perfect split found for each sub-problem does not guarantee the best overall solution.


\section{Evaluation Methods}\label{sec:evaluation}

\subsection{Power Grid Test Cases}

To evaluate the performance of the various solving strategies, we require test cases of the future residential power grid, meaning many decentralized electricity sources. Since the main focus of this work lies in the performance benchmarks of the quantum algorithms, entirely accurate real-world instances are not required. Following the steps presented here, we generate real-world inspired test cases with equal production and consumption volume.

\begin{enumerate}
    \item Choose a real-world power IEEE system test case provided by \texttt{pandapower}\footnote{\url{https://pandapower.readthedocs.io/en/v2.0.1/networks/power_system_test_cases.html}} as a basis for the grid topology.
    \item Rescale the network, such that the smallest power line is 50\,m in length.
    \item Replace all lines with residential power lines (NAYY 4x50 SE) and set the voltage level in the entire grid to 400\,V. Replace transformators with power lines.
    \item Randomly place producers and consumers on the busses of the network.
    \item Sample production and consumption data from shifted normal distributions: $p_i \sim \mathcal{N}(\mu=\pm 1\text{kW}, \sigma=0.5\text{kW})$.
    \item Ensure $\sum_i p_i = 0$ through scaling of production data.
    \item Run power flow calculation, as described in Sec.~\ref{sec:preprocessing}.
\end{enumerate}
For our experiments, we choose the available IEEE test cases between 39 nodes and 1888 nodes and generate five instances based on different seeds. Table~\ref{tab:cases} lists the number of producers for each instance and case considered\footnote{Additional naming for test cases is shortened in the following: E.g., \textsf{case89pegase} is called \textsf{case89}.}.
Fig.~\ref{fig:case39-example} shows \textsf{case39} as an example of a preprocessed power grid with indicated power flow.



\begin{table}[]
    \centering
    \caption{Overview of the test case settings}
\setlength{\tabcolsep}{4pt}
\begin{tabular}{l|rrrrr|rrrrr}
 & \multicolumn{5}{c}{$N_P$} & \multicolumn{5}{c}{$K$} \\
\hline\rule{0pt}{1.1\normalbaselineskip}%
\textsf{case39} & 20 & 20 & 19 & 19 & 18 & 9 & 6 & 9 & 7 & 6 \\
\textsf{case57} & 28 & 29 & 29 & 29 & 28 & 10 & 10 & 9 & 8 & 10 \\
\textsf{case89pegase} & 45 & 45 & 43 & 43 & 43 & 10 & 11 & 10 & 9 & 10 \\
\textsf{case118} & 58 & 58 & 58 & 58 & 60 & 14 & 11 & 11 & 9 & 11 \\
\textsf{case145} & 76 & 73 & 73 & 75 & 74 & 9 & 10 & 10 & 8 & 12 \\
\textsf{case\_illinois200} & 98 & 100 & 99 & 99 & 100 & 14 & 15 & 17 & 18 & 15 \\
\textsf{case300} & 148 & 147 & 149 & 153 & 147 & 20 & 18 & 19 & 17 & 19 \\
\textsf{case1354pegase} & 676 & 678 & 678 & 675 & 683 & -- & -- & -- & -- & -- \\
\textsf{case1888rte} & 942 & 949 & 939 & 939 & 941 & -- & -- & -- & -- & -- \\
\end{tabular}
\label{tab:cases}
\end{table}

\subsection{Parameters and Metrics}

We chose $\lambda = 0.1$ as the self-reliance penalty based on preliminary experiments across various grid topologies. This value ensures that communities maintain satisfactory self-sufficiency while also remaining intact. Too large self-reliance penalties would force communities to be ripped apart.


The only control parameter for the Leap cloud-based solvers is the runtime. Therefore, we restrict the runtime of all solvers per instance to the minimum required runtime of the Leap solver. Although the runtime is not comparable between local and cloud computing (and the hybrid machine runs unknown specifications), we can compare the scaling behavior in terms of solution quality regarding different instance sizes\footnote{Local classical optimizations are run on an Intel i7-1165G7 CPU @ 2.80GHz, 16.0 GB RAM.}.

The runtime is measured as the time it takes for the solver to produce a result; no model construction times are considered. Therefore, in the divisive approach, the sum of all sub-problem runtimes equals the overall runtime. The time for model creation can be further optimized and accelerated, so it is not included.

The parameters for the QUBO heuristics, SA and QSA, are set to 1000 Monte Carlo sweeps, and the sampling process is repeated until the specified timeout has been reached. 

The following benchmark protocol is used to compare the results.
For each case and instance, the optimization is repeated five times with different initial seeds for each solver. To compare the quality of a community structure $\mathcal{C}_{c,i,s,r}$ found by solver $s$ for case $c$, instance $i$, and repetition $r$, we calculate the solution energy (called SRCD energy) akin to the objective function~\eqref{eq:optimization}, i.e., modularity minus self-reliance violation: 
\begin{align}
      E(\mathcal{C}) = Q(\mathcal{C})-\lambda P(\mathcal{C}).
\end{align}
As a consequence, a higher SRCD energy is better.

Since we do not have access to the optimal solution of a case and instance, we need to make the solution energy comparable between instances. Therefore, we employ the best-found solution energy per instance as the reference point for the other solutions:
\begin{align}\label{eq:solution-quality}
e^{\text{rel}}_{c,i,s,r} = E(\mathcal{C}_{c,i,s,r}) \Bigm/ \left(\max_{s', r'} E(\mathcal{C}_{c,i,s',r'}) \right).
\end{align}
A relative solution quality $e^\text{rel}$ close to 1 means a very good solution in comparison to all other solutions found. It should be noted that $e^{\text{rel}}$ may be negative if the modularity value of the solution is low and self-reliance is violated.

To quantify self-sufficiency, we compute the average self-reliance violation, defined as follows
\begin{align}
    \frac{1}{|\mathcal{C}|} \sum_{C \in \mathcal{C}} \frac{\left|\sum_{i \in C}p_i\right|}{\sum_{i \in C}|p_i|}.
\end{align}
Similar to Ref.~\cite{zhao2019}, we define the Boundary Powerflow Factor (BPFF) as the normalized average power flow running between communities
\begin{align}
    \text{BPFF} = \frac{1}{B} \sum_{(i,j) \in E} \frac{A_{i,j}}{\max_{e \in E} A_e} (1 - \delta_{c_i, c_j}),
\end{align}
where $B$ is the number of lines between two communities and $(1 - \delta_{c_i, c_j})$ indicates whether line $(i,j)$ connects two different communities.

\subsection{SRCD Experiments}\label{sec:srcd-experiments}

\subsubsection{Single-Problem}

The single-problem approach from Eq.~\eqref{eq:srcd-qubo} has to be viewed separately from the divisive approach, as we can only set the timeout on a solver call level. We benchmark test cases up to 300 nodes with Gurobi, Leap CQM, SA, QSA, and Louvain.

Since the minimum runtime Leap requires for solving \textsf{case300} (6\,s) only differs slightly from the overall minimum runtime of the Leap CQM solver, being 5\,s, we set all solver timeouts for our experiments concerning the single-problem approach to 6\,s.

One major drawback of the single-problem approach is that~$K$---the number of communities---has to be known beforehand for the optimization, in contrast to the Louvain heuristic or the divisive approach. Therefore, before the optimization, we run the Louvain algorithm 100 times and pick the median number of communities found as~$K$. A too-large~$K$ is mathematically not a problem since communities can be kept empty. However, it increases complexity and, therefore, reduces solution quality. This has also been confirmed by preliminary experiments, where the presented strategy delivered the best results. The number of communities $K$ for each instance can be observed in Table~\ref{tab:cases}. Accordingly, the number of binary variables ($N \times K$) ranges from 234 in the smallest instance to 6000 in the largest.

\subsubsection{Divisive Algorithm Evaluation}
Results for the divisive approach are gathered with Gurobi, Leap BQM, SA, QSA, and D-Wave's Advantage 5.4 system (DW) on the cases (\textsf{case39}--\textsf{case300}). Here, the number of binary variables in the largest sub-problem equals the number of nodes in the network.
Sub-problems with $\leq 20$ nodes are solved with Gurobi. 
The minimum time limit of Leap BQM is 3\,s and gradually increases for problems larger than 1000 binary variables. The same timeout is set for the other solvers. Unlike the single-problem approach, the timeout is only set for the sub-problems. Consequently, two solvers may take different times to solve the SRCD problem because they solve a different number of sub-problems.

\subsubsection{First Split Evaluation}
The number of sub-problems solved in the divisive approach grows with the number of identifiable communities in the network ($\approx 2 K$) and, therefore, with the grid size. However, with each iteration, the problem size shrinks by, most likely, one-half. The remaining sub-problems then have the same structure. To isolate the performance of the solvers in the divisive approach, we focus on the first sub-problem of a network, i.e., the first split. Doing so allows us to investigate the performance of larger problems since the growing number of sub-problems does not need to be considered.

\section{Results}\label{sec:results}

\begin{figure}
    \centering
    \includegraphics[width=0.38\textwidth]{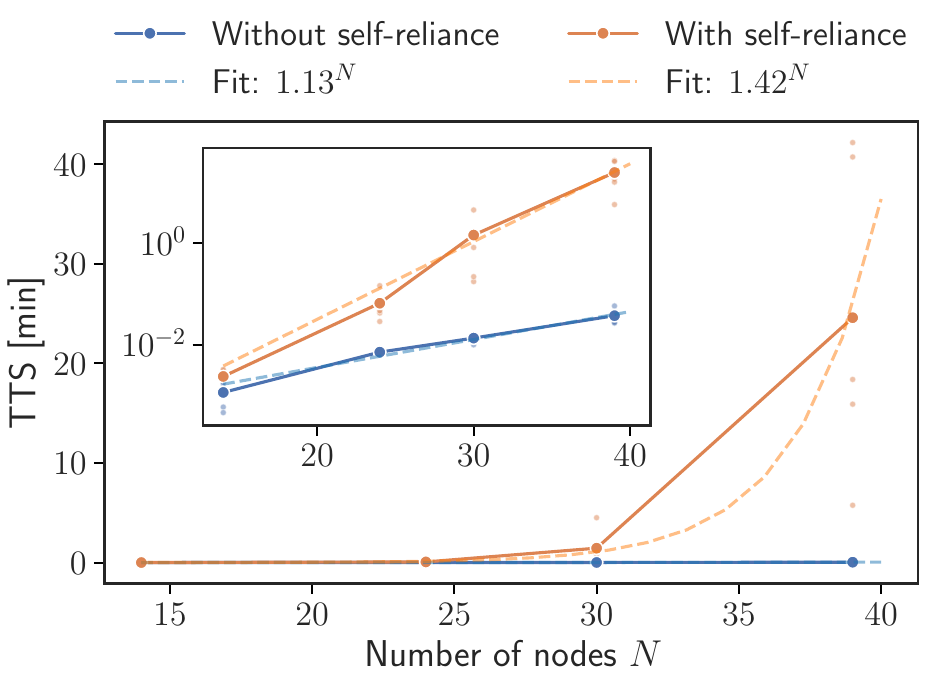}
    \caption{Gurobi time to solution (TTS) for five instances. A clear exponential growth in the required runtime is apparent. Results with self-reliance turned on and off are both shown. Evidently, self-reliance makes the problem much harder. The fitted curves highlight the exponential behavior.}
    \label{fig:gurobi-tts}
\end{figure}

We will first discuss the results of the two approaches separately due to different runtime settings. Nevertheless, in Sec.~\ref{sec:comparison-approaches}, we will still qualitatively compare the solutions yielded by both methods. 

\paragraph*{On Classical Hardness} Before we delve into the benchmarks, we briefly investigate the classical hardness of the considered problem in Fig.~\ref{fig:gurobi-tts}. Here, Gurobi solves the SRCD (with and without the self-reliance in use) until proven optimality, unlike the benchmark runs. The problem exhibits an exponential scaling behavior with the number of nodes in the network. With self-reliance in place, the scaling is even worse than plain community detection.

\subsection{Verification of the mathematical formulation}
\begin{figure}
    \centering
    \includegraphics[width=0.42\textwidth]{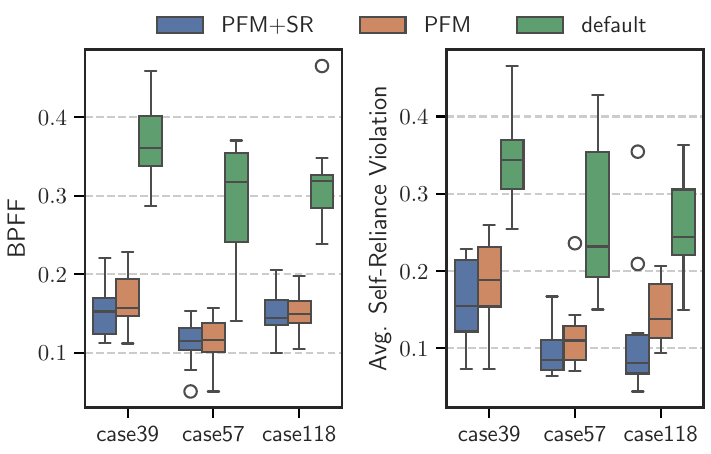}
    \caption{BPFF and avg. self-reliance violation of different CD methods for three power networks. \textsf{Default} refers to CD on just the network topology. \textsf{PFM+SR} and \textsf{PFM} label our formulation with and without the self-reliance penalty. Clearly, our method reduces power flow between communities and results in the most self-reliant communities.}
    \label{fig:verification}
\end{figure}

To verify that our formulation of the mathematical model also delivers on the intended goals, namely reducing power flow between communities and finding almost self-sufficient communities, we test the performance regarding the BPFF and avg. self-reliance violation in Fig.~\ref{fig:verification}. The results were obtained with the modified Louvain algorithm on ten instances of three power system test cases and show three different methods: our method with ($\lambda = 0.1$) and without the self-reliance penalty, and CD on just the network topology. It is apparent that our method with self-reliance penalty performs best in both metrics and, therefore, fulfills the intended purpose.

\subsection{Single-Problem Results}

\begin{figure}[t]
    \centering
\includegraphics[width=0.45\textwidth]{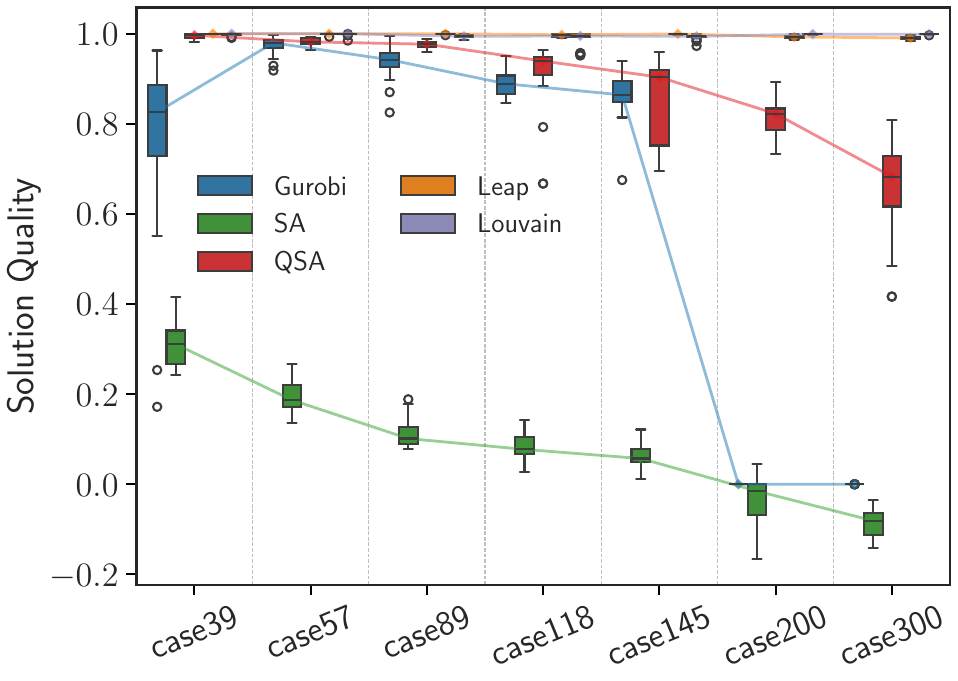}
    \caption{Solution qualities of the different solvers in the single-problem approach with 6\,s time limit. SA performs by far worst, followed by Gurobi, which produces solutions with acceptable quality for cases smaller than \textsf{case200}. However, in the larger cases, Gurobi fails to complete its preprocessing within the given time limit. QSA outperforms Gurobi but gradually decreases the larger the problem gets. Nevertheless, Leap and Louvain produce the best solutions; for a more detailed comparison see Fig.~\ref{fig:diff-leap-louvain}.}%
    \label{fig:sp-rel-eng}
\end{figure}

Fig.~\ref{fig:sp-rel-eng} shows the relative solution quality~\eqref{eq:solution-quality} of the five investigated solvers on seven IEEE power system test cases. Each box corresponds to 25 data points, consisting of five runs on each of the five instances per case. 

As apparent, the classical community detection heuristic Louvain and the cloud-based hybrid quantum solver Leap CQM produce the best results. A closer comparison can be seen in Fig.~\ref{fig:diff-leap-louvain} when we compare it to Leap's performance in the divisive approach. 
The QUBO heuristic, QSA, performs well in the smaller cases but continuously loses solution quality as the problems become larger. The classical branch-and-bound solver, Gurobi, is initially up to a similar course as QSA, but the solution quality suddenly diminishes to $e^{\text{rel}} = 0$ between \textsf{case145} and \textsf{case200}. This is due to Gurobi not finishing its preprocessing, consisting of solving a continuous relaxation of the original problem. Therefore, the branch-and-bound algorithm has not even started, and the returned solution assigns all prosumers to the same community. Finally, SA performs worst out of the considered solvers, most likely due to the one-hot constraint penalty term in Eq.~\eqref{eq:srcd-qubo} that requires a large penalty factor to be enforced. Furthermore, jumping from one feasible state to another requires at least two-bit flips. This significantly limits the exploration capabilities of the here-used single-bit-flip SA approach.

Of course, more specialized SA routines for one-hot encoding~\cite{kumagai2020}, as well as hyperparameter tuning, can be considered to improve the results. However, the authors choose to remain with the default implementation provided by D-Wave as this work focuses on the performance of the quantum solvers.

\subsection{Divisive Algorithm Results}
\begin{figure}[t]
    \centering
\includegraphics[width=0.42\textwidth]{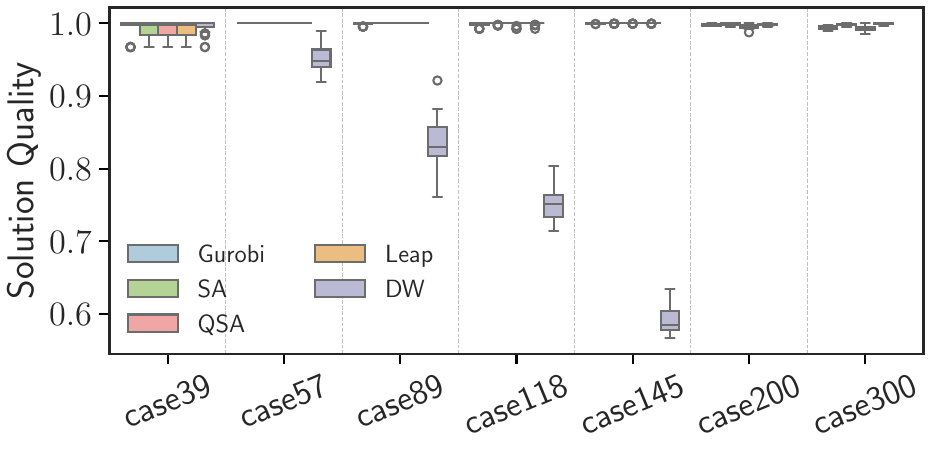}
    \caption{Solution qualities of the solvers in the divisive approach with $3\,\mathrm{s}$ time limit per sub-problem; all solvers, except for DW, produce good results. DW only really delivers comparable results for the smallest \textsf{case39} and performs more poorly as the size increases.}
    \label{fig:gcs-rel-energy}
\end{figure}

\begin{figure}[t]
    \centering
\includegraphics[width=0.42\textwidth]{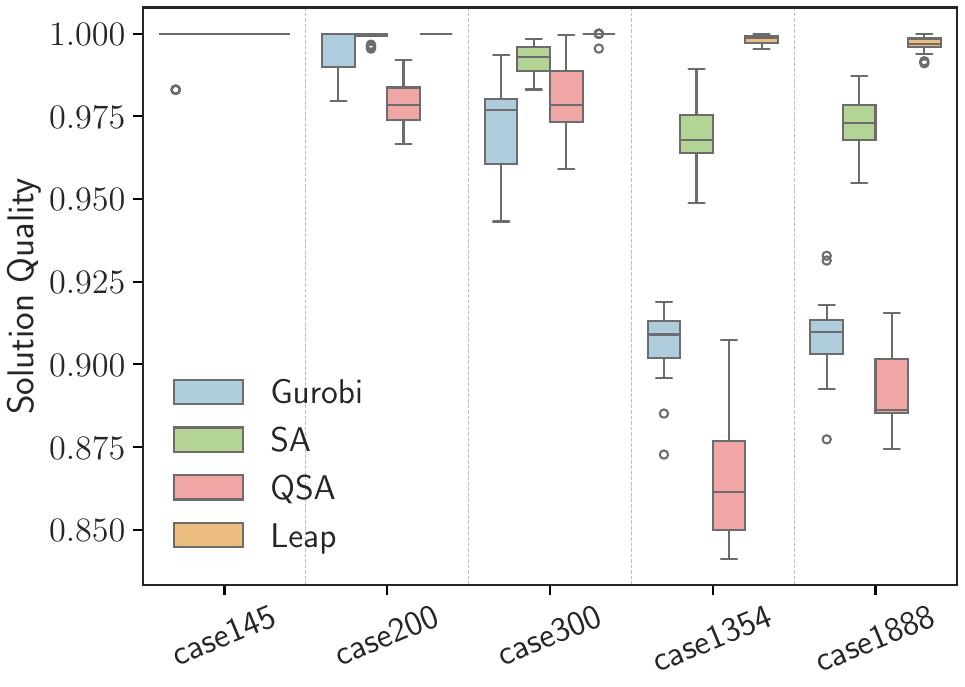}
    \caption{Solution qualities of the solvers in the first split of the divisive approach on larger problem instances. The performance tendencies indicated by the complete divisive run shown in Fig.~\ref{fig:gcs-rel-energy} are only accentuated here. Leap is followed by SA, Gurobi, and QSA. Time limit settings are $3\,\mathrm{s}$, except for \textsf{case1354} ($3.8\,\mathrm{s}$) and \textsf{case1888} ($5.0\,\mathrm{s}$), due to Leap's minimal runtime requirements.}
    \label{fig:first-split-rel-eng}
\end{figure}

The results for the divisive approach are displayed in Fig.~\ref{fig:gcs-rel-energy}. Here, we analyze the relative solution qualities of Gurobi, SA, QSA, Leap BQM, and DW as sub-problem QUBO solvers. 

Most notably, solutions in the divisive algorithm do not diverge as much as in the single-problem approach, except for DW.
While Leap performs consistently well, a substantial improvement for SA and Gurobi with larger instances can be observed compared to the single-problem approach. The lack of one-hot constraints in the sub-problem formulation of the divisive algorithm can most likely explain the vast performance improvement for SA. Except for DW and a couple of outliers, all algorithms consistently find the identical solution until \textsf{case145}, where we see some divergence between the solution quality for different sub-solvers.

The results of DW are significantly worse than the solutions of the classical and hybrid counterparts for cases other than \textsf{case39}. Because the solution quality drops so quickly, we run DW only until \textsf{case118}. Nonetheless, the results shown here are the best that have been obtained after extensive hyperparameter tuning: We utilize a dynamic chain strength that increases linearly with the sub-problem size: The \emph{chain strength factor}, i.e., the scaling of the automatically determined chain strengths, is calculated from the following expression $0.50 + 0.03 \times N$. The values in the formula have been estimated using linear regression on preliminary experiment data.
Based on the research of~\cite{marshall2019power} we include a $20 \upmu\mathrm{s}$ pause into our $60 \upmu\mathrm{s}$ annealing schedule of Eq.~(\ref{eq:time-evolution}) at the annealing fraction $s = 0.3$. Increasing the number of samples did not improve the results by much. Hence, we use a moderate 200 reads from the QPU. Overall, tuned hyperparameters reduce the gap between the best-found solution and the QPU solution substantially by 40\% on average. Additionally, the choice of embedding strategy (i.e., the algorithm that selects the chains) did not play a significant role. Therefore, we utilize clique embedding provided by D-Wave due to its fast wall-clock runtimes. A further improvement through approximation of the QUBO by pruning elements with small magnitude, similar to Ref.~\cite{sax2020}, has been attempted without success.

Furthermore, we investigate the classical and hybrid solvers' performance on the first split sub-problem in Fig. \ref{fig:first-split-rel-eng} for more substantial problem instances. Here, the gap in solution quality widens. While Leap performs persistently well and has the most compact variance between different runs, SA still manages to keep up even for huge instances. For Gurobi and QSA, we see a drop in relative performance between \textsf{case300} and \textsf{case1354}. This continues the divisive algorithm trend in Fig.~\ref{fig:gcs-rel-energy}. Furthermore, the effect is only accentuated by the problem size quadrupling between these cases. 

\subsection{Comparison of Approaches}\label{sec:comparison-approaches}
\begin{figure}[t]
    \centering
\includegraphics[width=0.42\textwidth]{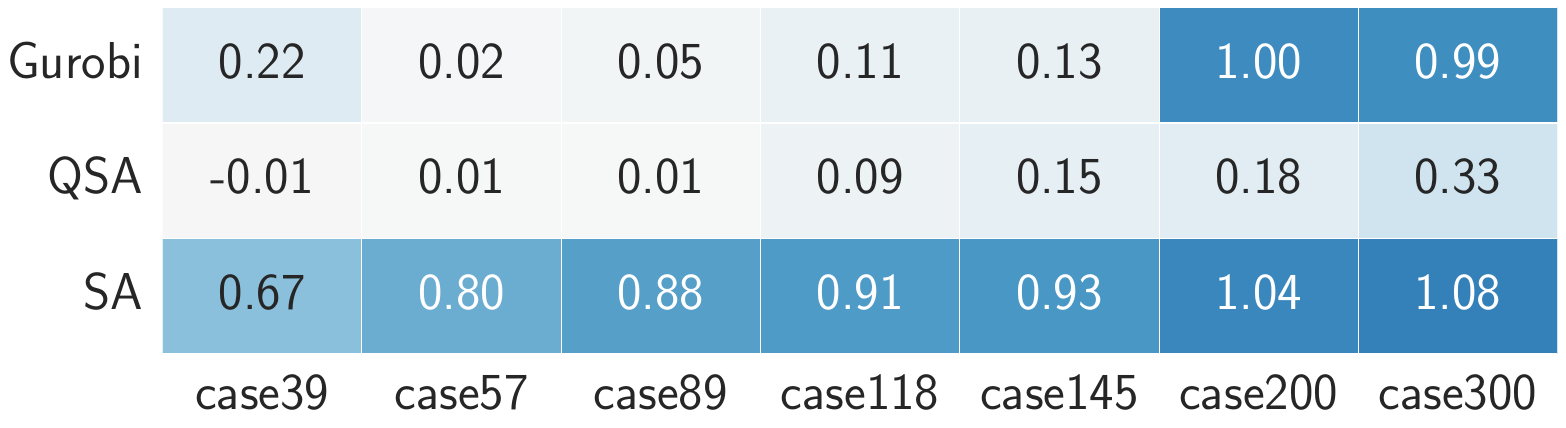} 
    \caption{Improvement of solution quality when using the divisive approach. Gurobi and QSA benefit for larger instances. SA shows a strong improvement for all cases and while it could not compete on the single-problem (Fig \ref{fig:sp-rel-eng}), with divisive it performs on par with the best solvers (Fig. \ref{fig:gcs-rel-energy}). However, the divisive approach requires more runtime than the single-problem approach as each sub-problem can run for $3\,\mathrm{s}$ compared to the single-problem runtime of $6\,\mathrm{s}$ total.}
    \label{fig:compare-approaches}
\end{figure}

\begin{figure}[t]
    \centering
\includegraphics[width=0.42\textwidth]{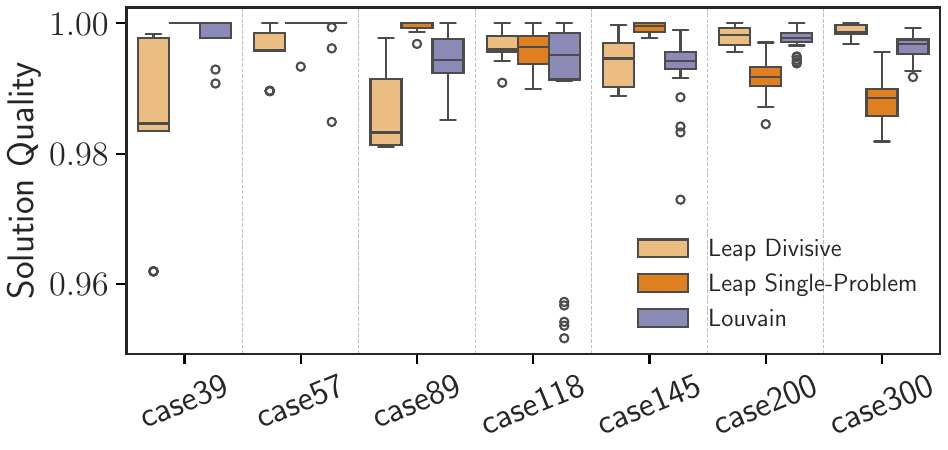}
    \caption{Comparison of single-problem Leap CQM, Louvain and divisive Leap BQM. In the smaller instances, Leap on single-problem approach produces the best results (up until \textsf{case145}). But in larger instances, Leap with the divisive approach wins.
    As expected by our definition, one solver will always achieve a relative solution quality of one. On average, one of the Leap solvers performs slightly better than Louvain for all case sizes.}
    \label{fig:diff-leap-louvain}
\end{figure}

As mentioned, the comparability between the single-problem formulation and the divisive approach is limited due to different runtime settings. However, it can still serve as an important guide for the relative performance of the different approaches. Furthermore, additional experiments have shown that even with an increased timeout set to the total runtime of the divisive algorithm, we do not see a great improvement for the single-problem versions of Gurobi and SA. Fig. \ref{fig:compare-approaches} shows that solvers utilizing the divisive approach can greatly improve their performance on larger instances. Here, SA benefits the most as the divisive sub-problem formulation does not require any constraints.

Finally, in Fig. \ref{fig:diff-leap-louvain}, the best solvers for each approach (Leap single-problem and Leap divisive) are compared with Louvain. Until \textsf{case145}, Leap on the single-problem outperforms Leap with the divisive algorithm since the single-problem formulation is more flexible and allows for discovering the optimal solution. The divisive approach, on the other hand, is greedy, and the optimal solution at every sub-problem will not guarantee the optimal overall solution. However, since the flexibility of the single-problem formulation comes with increased problem complexity, it becomes a disadvantage after \textsf{case145}, and the more restricted divisive approach returns better results. The classical heuristic Louvain consistently yields good results but, on average, not the best ones. It remains important to note that the single-problem approach with its 6\,s time limit is generally faster than the divisive approach, taking, for instance, $50$ to $70\,\mathrm{s}$ for Gurobi to finish \textsf{case300}. Auxillary experiments have shown that increasing the single-problem time limit to this runtime slightly improves the solution qualities for SA and Gurobi. However, the general pattern of Fig.~\ref{fig:sp-rel-eng} still remains.


\section{Conclusion}\label{sec:conclusion}

This research developed a mathematical problem formulation for finding self-reliant partitions of a power grid in terms of Self-Reliant Community Detection (SRCD). We demonstrated that solving the problem classically leads to an exponentially increasing runtime. Furthermore, we demonstrated the effectiveness of our problem formulations regarding the power flow between communities and the self-reliance of identified communities. Tailored to the problem formulation, we have presented two heuristics. The first one is based on a purely classical community detection heuristic called Louvain. The second is a hierarchical divisive algorithm, that iteratively splits sub-grids into two parts by solving QUBO sub-problems. The sub-problem QUBOs are significantly less demanding than the full-problem QUBO since the amount of binary variables is reduced and no one-hot constraints are necessary.

We benchmark the solution capabilities of various classical solvers (Gurobi, Louvain, SA, and QSA), cloud-based hybrid quantum-classical methods (Leap), and the direct usage of a QA processor (DW), both provided by D-Wave. The solution quality was measured on different IEEE grid test cases with multiple consumption and production data samples. Overall, Leap provided the most promising results when used on both the larger, more complex QUBOs and the sub-problem QUBOs of the divisive approach. The results indicate that the divisive approach outperforms the single-problem approach for more challenging instances. Nevertheless, our modified Louvain performs almost on par with the Leap solvers.

The remaining classical solvers performed significantly better in conjunction with the divisive algorithm than on the standalone problem. The performance difference ranges from similar (for QSA and Gurobi) to drastically better in the case of SA. Nevertheless, Gurobi fails to produce meaningful results within the given time constraints using the single-problem approach beyond problems with 200-node power grids.

Lastly, running the problem on the pure base-QPU demonstrated that the solution quality was considerably worse than that of its competitors and decreased with larger problem instances. Despite performing hyperparameter optimization for the problem at hand, the samples obtained from the QPU were only close to the optimal solution in the smallest instance \textsf{case39}.

The general problem identified here of self-reliant community detection could have real-world implications for renewable energy trading and sharing, and in the current search for quantum utility, identifying useful and computationally hard problems is vital to proving the usefulness of quantum computing algorithms to the real-world. Even though NISQ-era quantum devices are not yet up to the task of reliably solving these kinds of problems using only quantum computing resources, this work proves valuable in identifying another use case for quantum optimization and showcasing that quantum-enhanced (hybrid) solution techniques seem to offer competitive results in comparison to entirely classical methods.



\section*{Acknowledgements}
The authors would like to thank Kumar Ghosh for the helpful discussions throughout the course of this research.
This work was supported by the German Federal Ministry of Education and Research under the funding program ``Förderprogramm
Quantentechnologien – von den Grundlagen zum Markt'' (funding program quantum technologies — from basic research to market), project
Q-Grid, 13N16177.

\bibliographystyle{IEEEtran}
\bibliography{bstcontrol,references}

\end{document}


\maketitle


\section{QPU Hyperparameter Tuning}\label{app:hyperparameters}
To analyze and possibly improve the performance of the QPU in the divisive approach, we focus on individual bipartite splits, more precisely, the first split, and illustrate the process on \textsf{case89} as a representative. Like the main part, all experiments were conducted in five instances with five runs each. The initial results showed poor performance of the QPU, where the solution resembled a somewhat random assignment with only a few nodes connected by high-power flow reliably assigned to the same split. In comparison, SA provides clear visual separation between the two identified network sections.

Each sub-problem QUBO from Eq.~(17, main text) is dense and requires couplings between each pair of qubits.
Unfortunately, the D-Wave Advantage System only has a maximum connectivity of 15, making a native embedding of logical qubits connected to more than 15 others impossible. Thus, an embedding algorithm has to map a logical qubit to multiple physical qubits on the chip to ensure that the interactions between logical qubits are represented in their entirety. The set of physical qubits representing one logical qubit is called a \emph{chain}, and the coupling value between the individual chain elements is known as \emph{chain strength}, which is responsible for forcing them into a consistent state. When the qubits of the same chain are measured in different states, a so-called \emph{chain break} occurs. The state for the logical qubit will be, by default, determined via a majority vote. We observed that a higher proportion of chain breaks will decrease the quality of the solution. Therefore, one goal is to fine-tune the hyperparameters of the quantum annealing process to reduce chain breaks effectively. This can be done by increasing the chain strength or reducing the chain lengths to expose fewer qubits to conflicting biases (see Sec.~\ref{app:reduction}). Simply increasing the chain strength to an arbitrarily large value bears the risk of less flexible qubit states during the annealing process due to limited resolution in the physical implementation of the couplings. Therefore, a too-large coupling factor may have a decremental effect on solution quality.

\subsection{Hyperparamters}

\paragraph{Chain Strength} The chain strength is recommended to be close to the highest absolute value among the quadratic terms. By default, this is done by D-Wave's \texttt{uniform\_torque\_compensation}, compensating for conflicting biases on a chain. We scale the chain strength by a Chain Strength Factor (CSF) ranging from 1 to 5 and investigate its effects on solution quality. We also investigate whether the optimal chain strength differs among instances of the same case or the same-sized sub-problems of different cases. 

\paragraph{Annealing Time}
While not influencing the chain lengths, increasing the annealing time may also improve performance. D-Wave lets the user either define a custom annealing schedule or choose a constant annealing speed.
Refs.~\cite{marshall2019power, zielweski2022pause} show that custom schedules with annealing pauses might increase the chance of finding an optimal solution and also proposed a method to infer the optimal time and duration for the pause by empirically analyzing similar smaller sub-problems.

\paragraph{Number of Reads}
The annealing process is typically carried out multiple times, providing several samples during a solver call. When the best configurations are unlikely to be measured, a higher number of reads increases the chance of sampling them.

\paragraph{Embedding}
The embedding strategy is not a hyperparameter in a strict sense but will determine the algorithm used to map logical and physical qubits. D-Wave provides two strategies in their software repository: The deterministic \emph{clique} embedding and the \emph{heuristic} embedding through \texttt{minorminer}~\cite{choi2008minor}.
Intended for fully connected problems, clique embedding returns even-length chain lengths. It is accessed through \texttt{DWaveCliqueSampler}. In comparison, the heuristic embedding (accessed through \texttt{DWaveSampler}) has a higher variance in its chain lengths and, on average, longer chains, but it has the advantage of being more efficient on sparse problems. 

\subsection{Tuning Results}

\begin{figure*}
    \centering
    
    \subfloat[CSF\label{fig:csf-tuning}]{%
        \includegraphics[width=0.3\linewidth]{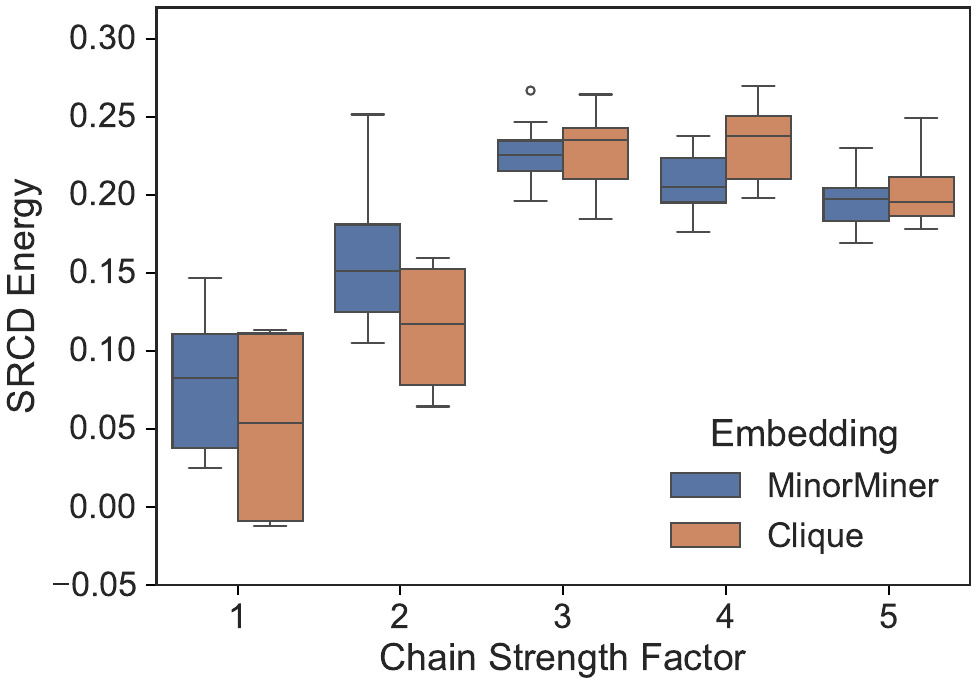}}
    \hfil
    \subfloat[Annealing Time\label{fig:tuning-annealing-time}]{%
        \includegraphics[width=0.3\linewidth]{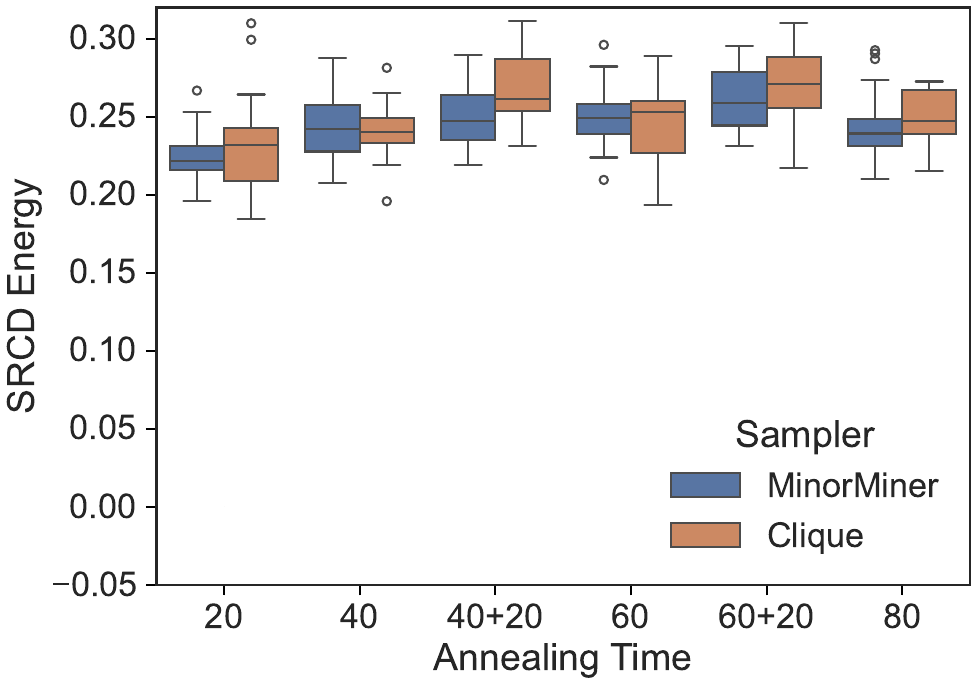}}
    \hfil
    \subfloat[Number of Reads\label{fig:tuning-reads}]{%
        \includegraphics[width=0.3\linewidth]{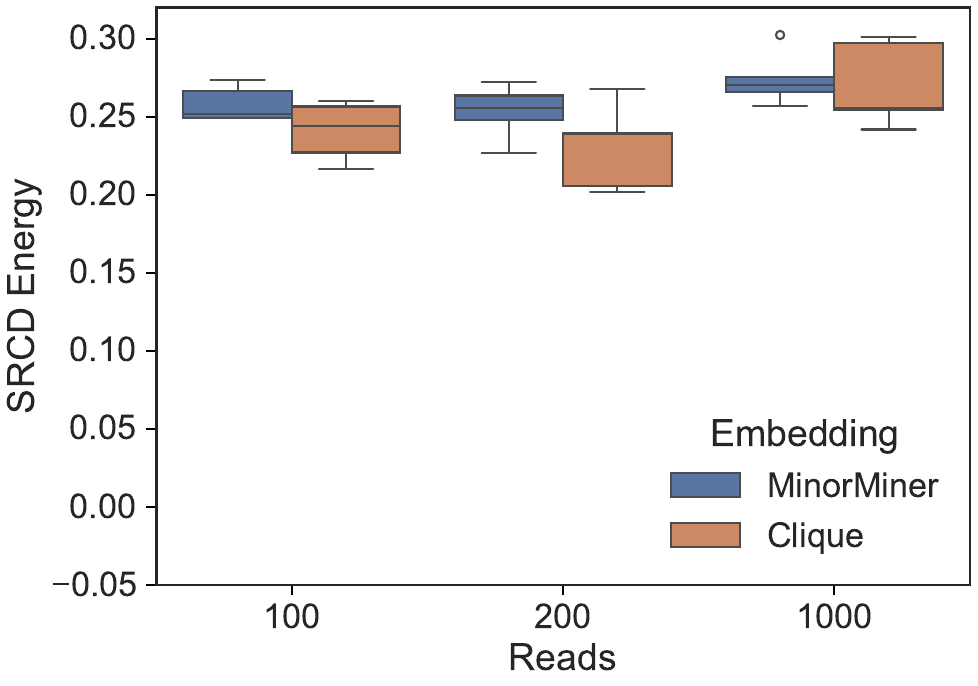}}
    \caption{The impact of successive parameter tuning on two embedding strategies on one instance of \textsf{case89}. Panel (a) shows that CSF is the dominant hyperparameter, significantly influencing solution quality. 
    In (b), solution quality concerning annealing time is depicted. An annealing time `40+20' refers to a $60 \upmu\mathrm{s}$ annealing process including a $20 \upmu\mathrm{s}$ pause at $s = 0.3$.
    As expected, a higher number of reads increases solution quality, see (c), since good results with low probability are more likely to be sampled. However, the added QPU cost is unjustified since the effect is small.}
    \label{fig:tuning-effects}
\end{figure*}

Experiments for hyperparameter tuning show a strong influence of the chain strength on the quality of the solution (see Fig.~\ref{fig:csf-tuning}). It is apparent that both the \texttt{DWaveCliqueSampler} and \texttt{DWaveSampler} perform best with a scaling factor of three in the representative \textsf{case89}. We found an almost linear relationship between the optimal chain strengths for \textsf{case39}, \textsf{case57}, \textsf{case89}, and \textsf{case118}, and sub-problems of their respective sizes derived from larger cases, which enables us to interpolate the optimal chain strength for any given problem size through the following expression
\begin{align*}
    \mathrm{CSF} = 0.496 + 0.031 \times N,
\end{align*}
which has been found through linear regression of the sampled data.
We apply this dynamic chain strength to the divisive algorithm since it presents sub-problems of many different sizes during the solving process. Overall, this improved the performance of the divisive method further compared to a static high chain strength and also the strength obtained by \texttt{uniform\_torque\_compensation}.

Fig.~\ref{fig:tuning-annealing-time} shows no significant difference among the constant annealing times. However, there is a slight improvement when a pause of $20~\upmu\mathrm{s}$ was inserted at a previously empirically determined annealing fraction $s = 0.3$.

The energies obtained from sampling resemble a normal distribution (see Fig. \ref{fig:tuning-num-samples}); hence, increasing the number of reads should also increase the chance of obtaining a better solution. Nevertheless, the observed results in Fig.~\ref{fig:tuning-reads} show only insignificant improvement between 100, 200, and even 1000 reads, indicating only a weak effect. To slightly increase the chance for better solutions without using a disproportionate amount of QPU time, we chose 200 reads as a compromise.

The clique embedding is particularly sensitive to chain strengths that are too low (see Fig.~\ref{fig:csf-tuning}), but performance did not differ from the minor embedding in the optimal case. No performance difference was observable for the different embedding strategies for the other hyperparameters we looked at. Nevertheless, clique embeddings are predetermined, which makes preprocessing fast compared to the \texttt{minorminer} heuristic and drastically reduces the wall-clock time.

Furthermore, we tested two Advantage systems: 5.4 in Germany and 4.1 in the US. But there was no difference worth mentioning between the systems. Since 5.4 is located in Germany (i.e., closer to the authors), the reduced network overhead caused faster overall execution times.

The general effect of the tuning parameters was consistent among different instances of the same power system test case.

\subsection{Summary}
We showed that tuning the hyperparameters is a crucial step when working directly with the QPU. Although finding the optimal hyperparameters takes additional time and resources, they are stable against changes in a power grid's power flow. Since electrical grids are not subject to constant significant changes in connections, hyperparameter tuning must only be performed once for a single power grid. We showed that chain strength was the main contributor to performance enhancement, with annealing time and the number of reads having minor roles. 

Overall, we did observe that tuning improved the solution quality drastically. In the full divisive algorithm, we improved the gap between the QPU solution and the optimal solution by 40\% on average. This is composed of 0\% in \textsf{case39} (already good results without tuning), about 40\% in \textsf{case57} and ca. 60\% in \textsf{case89}
and \textsf{case118}. The results indicate a clear trend of larger instances benefitting more from hyperparameter optimization.

Our analysis did not reveal any significant performance differences between the two embedding strategies or the two tested D-Wave Advantage Systems. However, using \texttt{DWaveCliqueSampler} on the Advantage system 5.4 significantly reduced the wall-clock time due to faster embedding and networking times. In summary, we have significantly improved the quantum annealer's performance by hyperparameter tuning.

\section{Problem reduction by approximation}\label{app:reduction}

While hyperparameter tuning and embedding strategies are two ways to reduce chain breaks without changing the problem, we also investigate how the problem formulation may be approximated in favor of better annealing performance. In particular, this is done by pruning QUBO interactions with small magnitude, lowering the required connectivity of the problem~\cite{sax2020}. 

\begin{figure*}
    \centering
    \subfloat[Improvement of solution space by tuning\label{fig:tuning-num-samples}]{%
        \includegraphics[width=0.3\textwidth]{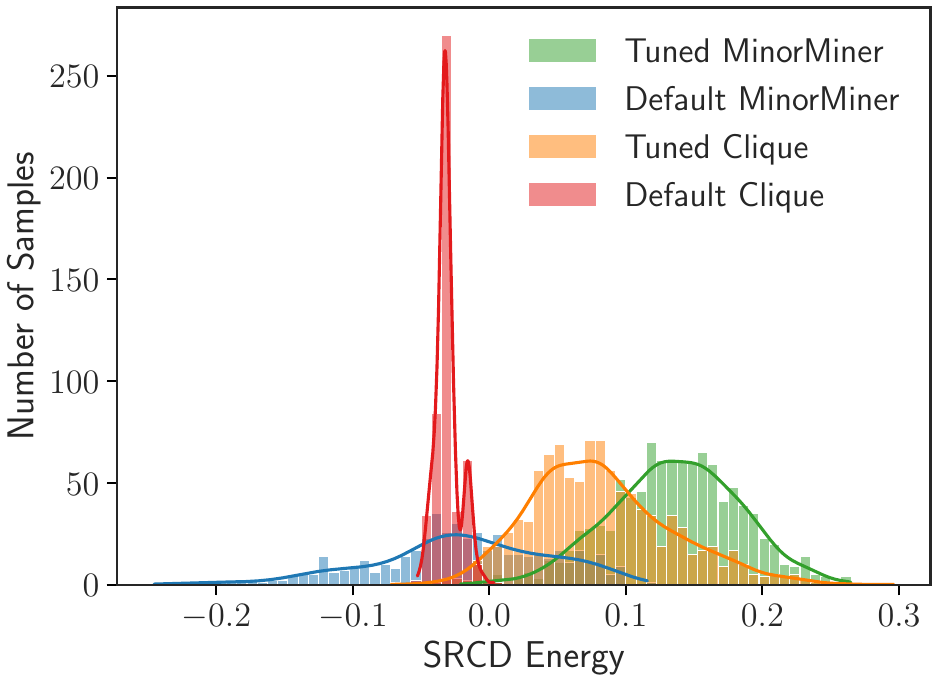}
    }
    \hfil
    \subfloat[Reduction effects for the QPU and SA\label{fig:reduction-qpu-sa}]{%
        \includegraphics[width=0.3\textwidth]{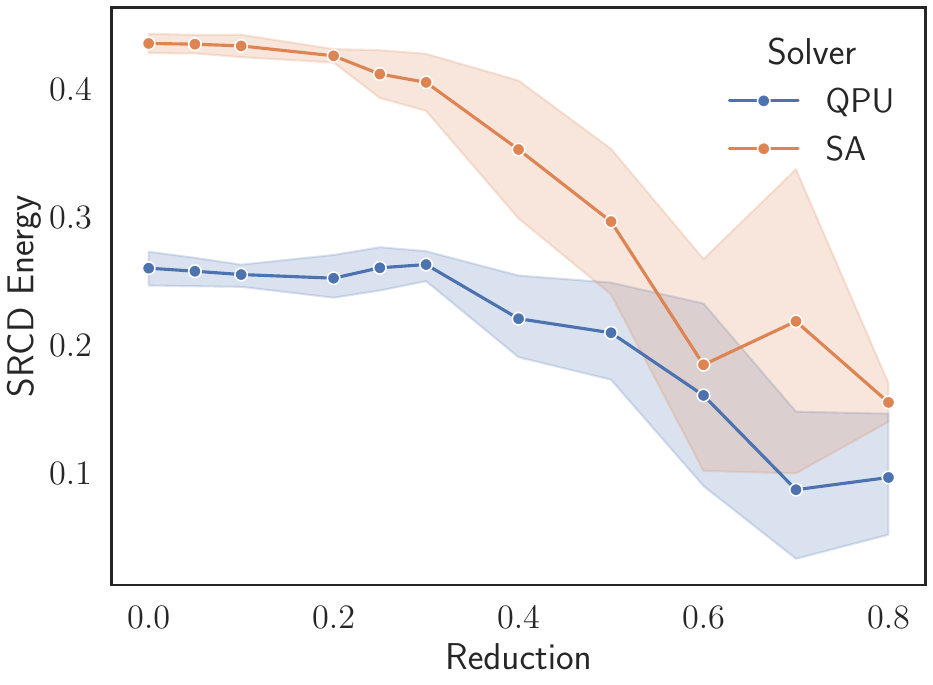}
    }
    \hfil
    \subfloat[Decrease of average chain lengths by reduction\label{fig:reduction-chains}]{%
        \includegraphics[width=0.3\textwidth]{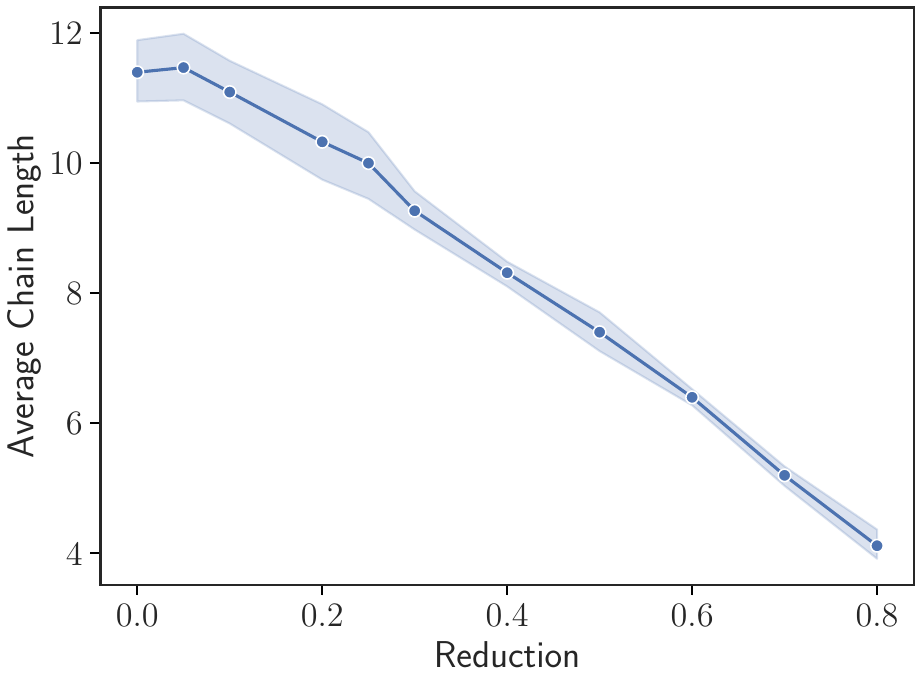}
    }

    \caption{Histogram of sample energies for both embedding strategies with and without tuning in Panel (a). Tuning improves the probability of sampling better solutions. Except for the untuned clique embedding, the distributions are close to normal distributions. The total number of samples for each distribution results from the number of reads multiplied by five runs. Panel~(b) compares the QPU results with SA when pruning the problem. Reducing quadratic terms approximates the problem and, therefore, also reduces the optimal solution quality. For the QPU, the approximation loss negates the gain from shorter chains up to a reduction ratio of 0.3, but no overall improvement can be observed. Chain lengths decrease with respect to the reduction ratio, as shown in Panel (c).}
    \label{fig:tuning-results}
\end{figure*}

The QUBO matrices of our problems included many close-to-zero quadratic terms that may not influence the solution but only lead to higher connectivity. Clique embedding does not benefit from reducing quadratic terms since it always assumes fully connected logical qubits. Hence, we only investigate the reduction on the \texttt{DWaveSampler}. Fig.~\ref{fig:reduction-qpu-sa} shows the solution quality with respect to the reduction fraction, where $0$ refers to the original problem and $0.8$ means that $80\%$ of the smallest magnitude couplings have been removed. Both SA and QPU exhibit almost constant solution quality until about 30\% of the entries have been pruned. For the QPU, one could argue that the solution quality slightly increases at a reduction of $0.3$. However, that effect is not really significant. As the reduction increases further, the solution quality drops as expected since the approximation alters the QUBO too much. The expected effect of the QPU's solution quality increasing as the problem complexity decreases could not be observed.
However, we can see an improvement regarding chain lengths as we saw an almost linear decrease in average chain length (see Fig.~\ref{fig:reduction-chains}). Naturally, shorter chains also positively affect the chain-break fraction, which describes the number of chains with contradicting qubit states after annealing.

Therefore, reducing the chain lengths by approximation did not help increase the quality of the QPU solution in the SRCD sub-problems.

\bibliographystyle{IEEEtran}
\bibliography{IEEEabrv,references}